\documentclass[apjl]{emulateapj}
\usepackage{times}

\newcommand{\XMM}{{\em XMM-Newton }}
\newcommand{\Ch}{{\em Chandra }}
\newcommand{\Su}{{\em Suzaku }}

\def\gappeq{\mathrel{ \rlap{\raise.5ex\hbox{$>$}}
                      {\lower.5ex\hbox{$\sim$}}  } }
\def\lappeq{\mathrel{ \rlap{\raise.5ex\hbox{$<$}}
                      {\lower.5ex\hbox{$\sim$}}  } }

\shorttitle{The Suzaku X-Ray View of NGC\,6251}
\shortauthors{EVANS ET AL.}

\begin{document}

\title{The Suzaku View of the Disk-Jet Connection in the Low Excitation Radio Galaxy NGC 6251}
\author{D.~A.~Evans\altaffilmark{1,2}, A.~C.~Summers\altaffilmark{2}, M.~J.~Hardcastle\altaffilmark{3}, R.~P.~Kraft\altaffilmark{1}, P.~Gandhi\altaffilmark{4}, J.~H.~Croston\altaffilmark{5}, J.~C.~Lee\altaffilmark{1}}
\altaffiltext{1}{Harvard-Smithsonian Center for Astrophysics, 60 Garden Street, Cambridge, MA 02138}
\altaffiltext{2}{Elon University, 100 Campus Drive, Elon, NC 27244}
\altaffiltext{3}{School of Physics, Astronomy \& Mathematics, University of Hertfordshire, College Lane, Hatfield AL10 9AB, UK}
\altaffiltext{4}{Institute of Space and Astronautical Science (ISAS), Japan Aerospace Exploration Agency, 3-1-1 Yoshinodai, Chuo-ku, Sagamihara 252-5210, Japan}
\altaffiltext{5}{School of Physics and Astronomy, University of Southampton, Southampton, SO17 1BJ, UK}

\begin{abstract}

We present results from an 87-ks \Su observation of the canonical low-excitation radio galaxy (LERG) NGC\,6251. We have previously suggested that LERGs violate conventional AGN unification schemes: they may lack an obscuring torus and are likely to accrete in a radiatively inefficient manner, with almost all of the energy released by the accretion process being channeled into powerful jets. We model the 0.5--20 keV \Su spectrum with a single power law of photon index $\Gamma=1.82^{+0.04}_{-0.05}$, together with two collisionally ionized plasma models whose parameters are consistent with the known galaxy- and group-scale thermal emission. Our observations confirm that there are no signatures of obscured, accretion-related X-ray emission in NGC 6251, and we show that the luminosity of any such component must be substantially sub-Eddington in nature.

\end{abstract}

\keywords{galaxies: active -- galaxies: jets -- galaxies: individual (NGC\,6251) -- X-rays: galaxies}

\section{INTRODUCTION}
\label{intro}

The origin of relativistic jets in active galactic nuclei (AGN) is a key unsolved problem in extragalactic astrophysics. While 90\% of all AGN (Seyfert galaxies and radio-quiet quasars) show little or no jet emission, the remaining 10\% (the radio-loud AGN and radio-loud quasars) launch powerful twin jets of particles from their cores. Since jets transport a significant fraction of the energy liberated during the accretion process, sometimes out to $\sim$Mpc distances, understanding how they are produced is critical to a complete picture of accretion and feedback in AGN.

X-ray observations of the nuclei of radio-loud AGN are particularly useful for establishing the connection between the accretion flow, black hole, and jet. Since the launch of {\it Chandra} and {\it XMM-Newton}, several groups have performed spectroscopic studies of the nuclei of 3CRR radio galaxies (e.g., \citealt{evans06,bal06,hec07,hec09}). These efforts have found a dichotomy in the nuclear X-ray spectra of radio-loud AGN, related to the optical emission line classifications, which may be related to their differing roles of hot and cold gas accretion (\citealt{hec07}), or possibly the spin configuration of the black hole itself (\citealt{ges10}).

High-excitation radio galaxies (HERGs), those with prominent emission lines in their optical spectra, have X-ray spectra that are consistent with unification models. HERGs have standard, cool luminous accretion disks and strong (yet typically unbroadened) Fe K$\alpha$ lines. The nuclear continuum is heavily obscured in the X-ray by cold gas with columns of $\sim$$10^{23}$ cm$^{-2}$ (\citealt{evans06}) when the source AGN is viewed close to edge-on with respect to the observer, and are largely unobscured when oriented closer to the line of sight. These sources also have an additional component of unabsorbed emission that is associated with the pc-scale jet. HERGs tend to occupy gas-poor environments (e.g., \citealt{kra07}) and their host galaxies lie in the `green valley' of the galaxy color-mass diagram (\citealt{smo09}).

Low-excitation radio galaxies (LERGs), on the other hand, the population of radio-loud AGN with little or no nuclear optical line emission, {\it seem to lack any of the features required by standard AGN unification models}. We have previously argued that their X-ray emission is dominated by radiation from a parsec-scale jet, that they have no signatures of standard, cold accretion disks, and that they may show no evidence at all for an obscuring torus (\citealt{ogle06,hec09}). We have suggested that accretion in LERGs takes place in a radiatively inefficient manner, with almost all the available energy from accretion being channeled into jets. LERG host galaxies lie in the red sequence on the color-mass diagram (\citealt{smo09}) and are often associated with group- or cluster-scale hot-gas environments (e.g., \citealt{tas08})

Constraining the properties of the accretion flow in LERGs with deep X-ray observations would provide a strong test of our hypothesis that the X-ray emission is dominated by a jet with no signatures of cold-gas accretion. However, existing observations of LERGs have been restricted to relatively poor signal-to-noise \Ch and \XMM data, while the $>$10~keV band where the Compton reflection hump could lie is completely inaccessible to those observatories. In order to address these issues, we obtained an 87-ks \Su observation of NGC\,6251 ($z$=0.0247), the X-ray brightest low-excitation radio galaxy in the 3CRR catalog. We use the XIS and HXD to search for signatures that are related to a standard, cool accretion disk that is surrounded by a torus: (1) a $\sim$6.4~keV Fe~K$\alpha$ line, (2) heavily absorbed continuum emission, and (3) the $>$10~keV Compton reflection hump (\citealt{geo91}). Detection of any of these features would immediately invalidate our model.

Previous attempts to understand the origin of X-ray emission in NGC\,6251 have yielded conflicting results, making our \Su observation particularly useful. For example, \cite{tur97} (based on {\it ASCA} data) and \cite{gli04} (based on \XMM data) claimed a detection of a high-equivalent width Fe~K$\alpha$ line, suggestive of a dominant contribution from a luminous accretion disk. However, reanalysis of the \XMM spectra by \cite{evans05} instead showed no evidence of an Fe~K$\alpha$ line, which favors a jet-dominated scenario. Furthermore, the double-peaked SED of NGC\,6251 measured by, e.g., \cite{ho99}, \cite{gua03}, \cite{chi03}, \cite{evans05}, and \cite{mig11} led those authors to conclude that the high-energy emission is synchrotron self-Compton (SSC) emission from a jet. Finally, analysis of the spectral variability from a long-term monitoring campaign with {\it RXTE} (\citealt{gli08}) again suggests that the jet, rather than the accretion flow, dominates the X-ray emission. Nonetheless, only the combination of high effective area and simultaneous soft and hard band offered by \Su can resolve once and for all the issues surrounding the Fe~K$\alpha$ line, as well as place strong limits on the luminosity of the accretion flow.

This paper is organized as follows. Section~\ref{observations} provides a summary of our observations and a description of their reduction. In Section~\ref{analysis}, we 
present the results of our spectral fitting to the XIS and HXD data. In Section~\ref{fek}, we show that an Fe~K$\alpha$ line is not detected, while in Section~\ref{accretion} we discuss whether or not a buried AGN is present in NGC\,6251. We end with our conclusions in Section~\ref{conclusions}.

\section{Observations And Data Reduction}
\label{observations}

We observed NGC\,6251 with \Su on 2010 December 02 (OBSID 705039010) for a nominal exposure of 87~ks. Both the X-ray Imaging Spectrometer (XIS) and Hard X-ray Detector (HXD) were operated in their normal modes. The source was positioned at the nominal aimpoint of the XIS instrument. The data were processed using v. 2.5.16.29 of the \Su processing pipeline, which includes the latest Charge Transfer Inefficiency (CTI) correction applied for the XIS. We used the standard cleaned events files, which are screened to remove periods during which the satellite passed through the South Atlantic Anomaly (SAA), had a pointing direction $<$5$^\circ$ above the Earth, or had Earth day-time elevation angles $<$20$^\circ$. We describe our analysis of the XIS and HXD data below.

\subsection{XIS}

We used data from the two operational front-illuminated (FI) CCDs (XIS0 and XIS3), together with the back-illuminated XIS1 detector. The three XIS CCDs were operated with a frame time of 8~s. For our analysis, we used data taken in the 3$\times$3 and 5$\times$5 edit modes. We selected only events corresponding to grades 0, 2, 3, 4, and 6, and removed hot and flickering pixels with the {\sc cleansis} tool.

We extracted the spectrum of NGC\,6251 from the XIS CCDs using a source-centered circle of radius 260$''$ (250 physical pixels), with background sampled from an adjacent region free from any unrelated sources, as well as the $^{55}$Fe calibration sources at the corners of each detector. We generated response matrix files (RMFs) for each detector using v. 2009-02-28 of the {\sc xisrmfgen} software, and ancillary response files (ARFs) using v. 2009-01-08 of the {\sc xissimarfgen} software. 

The net exposure times and count rates for the three CCDs are shown in Table~\ref{obslog}. We co-added the two FI spectra using the {\sc addascaspec} program, and grouped the resulting spectrum and that of the XIS1 detector to a minimum of 50 counts per bin in order to use $\chi^2$ statistics. We restricted the energy range for our spectral fitting to 0.5--10 keV.

\subsection{HXD}

We extracted the source spectrum from the HXD/PIN detector, using the cleaned PIN events files described above. The source was not detected with the GSO instrument. We used the {\sc hxdpinxbpi} script, which generates the PIN non X-ray background (NXB) spectrum from the latest `tuned' time-dependent instrumental background event file provided by the \Su Guest Observer Facility. The script also extracts the source and background spectra using a common Good Time Interval (GTI) criterion, creates a Cosmic X-ray Background (CXB) spectrum, and adds the CXB and NXB spectra together. We adopted the default binning criterion of 20 counts per channel. NGC\,6251 is detected at energies between 16 and 20 keV, and so we restrict our subsequent spectral analysis to this range.

\begin{figure}
\includegraphics[height=8cm,angle=270]{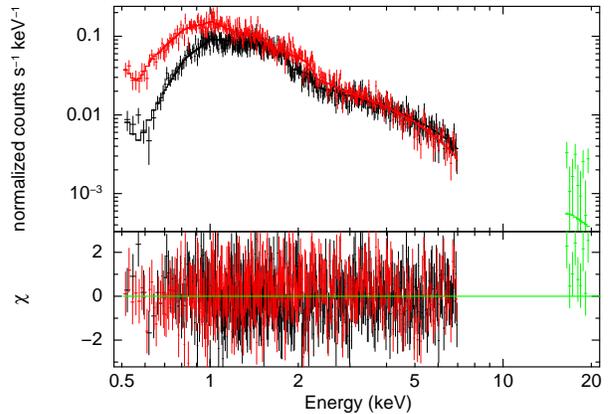}
\caption{\Su XIS FI {\it (black)}, XIS BI {\it (red)}, and HXD PIN {\it (green)} best-fitting spectra and residuals. The model fit is the sum of a power law with slight excess absorption (consistent with the known dusty disk), and two collisionally ionized plasma components {\bf (Model IV)}.}
\label{model4}
\end{figure}

\section{Spectral Analysis}
\label{analysis}

In all spectral fits, we linked the parameters of the model components across the \Su XIS FI, XIS BI, and HXD PIN spectra. We tied the normalizations of the FI and BI XIS spectra together, but applied a constant cross-normalization factor of 1.16 for the PIN spectrum relative to the XIS as described in the \Su Data Reduction Guide\footnote{http://heasarc.gsfc.nasa.gov/docs/suzaku/analysis/abc/}. We performed our spectral fitting using v.12.6.0 of the {\sc XSPEC} spectral fitting package. All results presented here use a cosmology in which $\Omega_{ m, 0}$ = 0.3, $\Omega_{\rm \Lambda, 0}$ = 0.7, and H$_0$ = 70 km s$^{-1}$ Mpc$^{-1}$. At the redshift of NGC\,6251, $z$=0.02471, the luminosity distance is 267.3~Mpc. Errors quoted are 90 per cent confidence for one parameter of interest (i.e., $\chi^2_{\rm min}$ + 2.7), unless otherwise stated. All spectral fits include the Galactic absorption to NGC\,6251 of $N_{\rm H, Gal}=1.69\times10^{20}$ cm$^{-2}$ (\citealt{dic90}).

We initially attempted to model the spectrum using a single, unabsorbed power law, but this resulted in a poor fit ($\chi^2$=1034 for 832 dof) and noticeable residuals around 1~keV ({\bf Model I}) that suggested the need for a thermal component. Following the methods of \cite{gua03}, \cite{gli04}, and \cite{evans05}, we subsequently added a collisionally ionized plasma ({\sc Apec}) component with its abundance fixed at 0.35 of solar (similar to \citealt{evans05}), which resulted in a substantial improvement to the fit ($\Delta\chi^2$=134 for 2 additional parameters), but still gave significant residuals below 1 keV ({\bf Model II}). Next, we added a second {\sc Apec} component to the model, which improved the fit to $\chi^2$=871 for 828 dof ({\bf Model III}). Finally, owing to this model's overprediction of flux below 0.7 keV, we added neutral intrinsic absorption at the redshift of NGC\,6251 to the power law component ({\bf Model IV}). This resulted in a substantial improvement to the fit ($\Delta\chi^2$=33 for 1 additional parameter). The best-fitting spectrum and model are shown in Figure~\ref{model4}. The addition of additional components, such as heavily obscured emission, failed to provide a significant improvement to the fitting statistic, and so we adopt Model IV as our best fit in the following discussion.

The measured intrinsic absorption is $N_{\rm H}$=$(7.68^{+1.26}_{-2.18})\times10^{20}$~cm$^{-2}$. The detection of modest absorption at the redshift of NGC\,6251 is not surprising, given the known dusty disk in the host galaxy. Indeed, the visual extinction of $A_{\rm V}= 0.61\pm0.12$~mag (\citealt{fer99}) corresponds to a neutral hydrogen column density of $(1.3\pm0.3)\times10^{21}$~cm$^{-2}$, which is close to the measured value from the {\it Suzaku} spectrum. The power law photon index of $\Gamma$=$1.82^{+0.04}_{-0.05}$ is consistent with the range of values from other X-ray observatories presented by \cite{evans05}. The 2-10 keV unabsorbed luminosity of the power law is ($2.78^{+0.17}_{-0.24})\times10^{42}$ ergs~s$^{-1}$, which again is within the range of historical values discussed by \cite{evans05}. We also detect two collisionally ionized plasmas, with temperatures $0.80\pm0.15$ keV and $kT$=$1.42^{+0.45}_{-0.20}$ keV. Their temperatures are consistent with the extended emission detected by \XMM on scales of tens and hundreds of kpc, respectively (\citealt{sam04,evans05}). Furthermore, the fraction of the integrated luminosity of the beta profile determined by \cite{evans05} within our 260$''$ \Su extraction region is approximately 60\%. This corresponds to a bolometric luminosity of $4.4\times10^{41}$~ergs~s$^{-1}$, which is consistent with the luminosity we measure with {\it Suzaku}.

\begin{figure}
\includegraphics[width=8cm,angle=0]{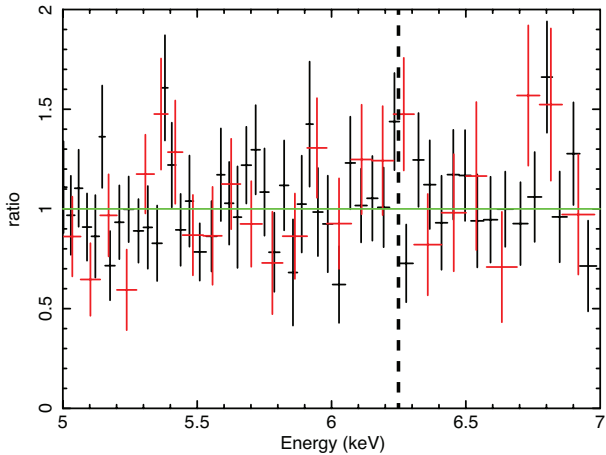}
\caption{\Su FI {\it (black)} and BI {\it (red)} data/model ratio in the energy range 5--7 keV. Also marked is the position where a neutral Fe~K$\alpha$ line would lie (dashed line).}
\label{fekzoom}
\end{figure}

\section{No Evidence of Iron K$\alpha$ Emission}
\label{fek}

A key goal of our \Su observation was to search for Fe~K$\alpha$ emission, which would indicate the presence of reflection from an accretion disk, and/or circumnuclear torus. Figure~\ref{fekzoom} shows the XIS FI and BI data/model ratios of NGC\,6251 in the energy range 5--7 keV. It is immediately evident that no Fe~K$\alpha$ line is detected with {\it Suzaku}. Nonetheless, we added to our best-fitting spectral model a Gaussian line with its energy fixed at 6.4~keV and its width fixed at 50~eV (below the instrumental resolution of XIS). This resulted in a small improvement to the fit ($\Delta\chi^2$=1.38 for 1 additional parameter), with a probability of 24\% that the improvement is due to statistical fluctuations alone. The lower limit to the normalization of the Gaussian is consistent with zero, while the upper limit to the equivalent width (with respect to the overall continuum) is $\sim$100~eV. We therefore conclude that there is no substantial evidence for neutral Fe~K$\alpha$ emission, although we caution that we cannot rule out its presence entirely.

\begin{figure}
\includegraphics[height=11cm,angle=0]{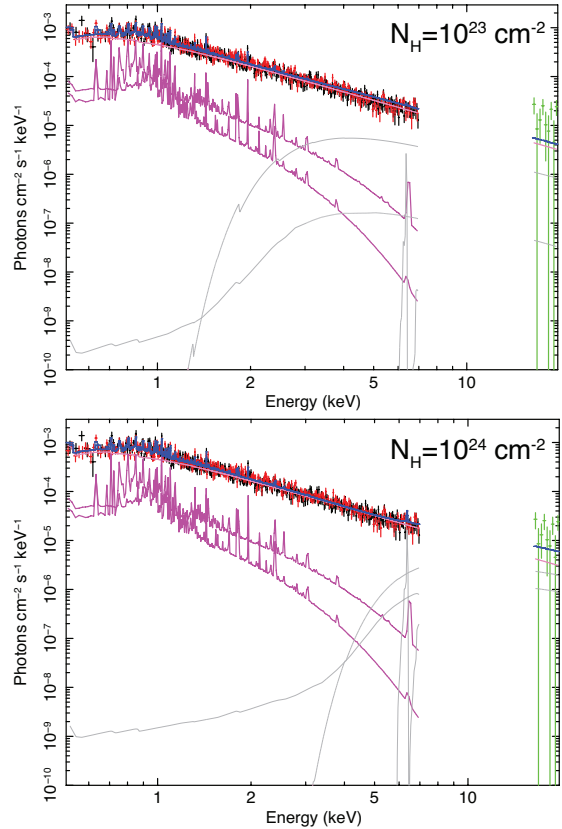}
\caption{\Su XIS FI {\it (black)}, XIS BI {\it (red)}, and HXD PIN {\it (green)} unfolded spectra. The overall model fit shown {\it (blue)} consists of an absorbed power law {\it (pink)} and two thermal components {\it (maroon)}, plus the direct, reflected, and fluorescent components of MYTorus {\it (gray)}. We have assumed a column density of $10^{23}$ cm$^{-2}$ {\it (upper panel)} and $10^{24}$ cm$^{-2}$ {\it (lower panel)}. The absorbed power law had a best-fit normalization of 0, but for illustrative purposes we show its 3$\sigma$ upper limit. In both cases, we assumed that there were no systematic uncertainties in the PIN background.}
\label{mytorus_try6and7}
\end{figure}

\section{Is a Buried AGN Present in NGC\,6251?}
\label{accretion}

We next used the entire \Su bandpass to search for the presence of any signatures related to an accretion flow and torus. We assumed that, in addition to the dominant power-law component (which \citealt{evans05}  interpreted to be associated with a jet), there exists a ``buried'', highly obscured AGN. We added to our best-fitting spectral description the MYTorus table models (\citealt{my09}), a fully self-consistent treatment of reflection and absorption from a circumnuclear torus. Specifically, MYTorus includes components that represent the (absorbed) zero-order continuum, the scattered continuum, and resulting neutral fluorescent lines. We assumed (1) the primary (accretion-related) power law has a photon index of 1.7 (consistent with the mean photon index of $z<0.1$ 3CRR sources found by \citealt{evans06}); and (2) it is attenuated by equatorial column densities of $10^{23}$~cm$^{-2}$ (consistent with HERGs measured by \citealt{evans06} and \citealt{bal06}) or $10^{24}$~cm$^{-2}$  (to test the Compton-thick case). The choice of inclination angle is problematic: \cite{jon02} used the jet--counterjet brightness ratio to conclude that the angle to the line of sight, $\theta$, of the jet is $\lappeq$45$^{\circ}$, but this is different to that subtended by the dusty disk (76$^{\circ}$) (\citealt{fer99}). For illustrative purposes, we adopt an angle of 90$^{\circ}$ (i.e., an edge-on geometry) and note that our results are essentially unaffected by the choice of angle until $\theta<60^{\circ}$, after which the reprocessor adopted by \cite{my09} no longer intersects our line of sight.

The XIS and HXD spectra and models for the $10^{23}$~cm$^{-2}$ and $10^{24}$~cm$^{-2}$ cases are shown in Figure~\ref{mytorus_try6and7}. The best-fitting accretion-related power law normalizations were consistent with zero, with 3$\sigma$ upper limits to the 2--10 keV unabsorbed luminosity of $7\times10^{41}$ ergs~s$^{-1}$ and $4\times10^{42}$ ergs~s$^{-1}$, respectively. Assuming a black hole mass of $6\times10^8$~M$_{\rm \odot}$ (\citealt{fer99}), we find that any obscured, accretion-related emission is highly sub-Eddington ($L_{\rm X}/L_{\rm Edd}\lappeq10^{-5}$ for $N_{\rm H}$=$10^{23}$~cm$^{-2}$; $L_{\rm X}/L_{\rm Edd}\lappeq5\times10^{-5}$ for $N_{\rm H}$=$10^{24}$~cm$^{-2}$). The upper limit to the 2-10 keV luminosity in both cases is consistent with the relationship  between X-ray luminosity and 15$\mu$m luminosity established by \cite{hec09}.

As an additional test, we considered an extreme Compton thick obscurer ($N_{\rm H}$=$10^{25}$~cm$^{-2}$). Again, this failed to provide a significant improvement to the fit. The upper limit to the 2--10 keV unabsorbed luminosity in this case is $2\times10^{44}$ ergs~s$^{-1}$. However, we note that (1) the peak of the Compton reflection hump lies outside of the usable energy range of the PIN data, meaning that the principal feature used to determine the strength of the accretion-related emission is inaccessible, and (2) a luminosity of $2\times10^{44}$ ergs~s$^{-1}$ would be a 6$\sigma$ outlier from the \cite{hec09} $L_{\rm X}$-$L_{\rm 15\mu m}$ relationship. Observations with {\it NuSTAR} and {\it Astro-H} would provide further constraints.

Finally, given the poor signal-to-noise of our PIN data, it is important to address the known systematic uncertainties of the \Su background, which are of order 3\% at the 1-sigma level (\citealt{fuk09}). In the above analysis, we noticed several points in the PIN band that appear to lie above the model. However, increasing the background by 3\% (set using the cornorm parameter in {\sc Xspec}) removes any discrepancy, ruling out any evidence of a hard excess given the PIN systematics.

\section{Conclusions}
\label{conclusions}

We have presented results from a new 87-ks \Su observation of the prototypical low-excitation radio galaxy NGC\,6251. We have shown the following:
\\
\begin{enumerate}
\item We model the 0.5--20 keV \Su spectrum with a single power law of photon index $\Gamma=1.82^{+0.04}_{-0.05}$, together with two collisionally ionized plasma models whose parameters are consistent with the known galaxy- and group-scale thermal emission.
\item There is no significant evidence in the X-ray for obscured, accretion-related emission. However, we cannot rule out accretion-flow luminosities of $<$$7\times10^{41}$ ergs~s$^{-1}$ ergs~s$^{-1}$ and $4\times10^{42}$ ergs~s$^{-1}$ if they are obscured by columns of $10^{23}$~cm$^{-2}$ or $10^{24}$~cm$^{-2}$, respectively. These luminosities are consistent with those predicted by the $L_{\rm X}$-$L_{\rm 15\mu m}$ relationship established by \cite{hec09}. Both luminosities are highly sub-Eddington ($L_{\rm X}/L_{\rm Edd} \sim$ $10^{-5}$).
\item NGC\,6251 is thus consistent with our predictions for low-excitation radio galaxies: they have radiatively inefficient accretion flows, show no evidence in the X-ray for circumnuclear tori, and their X-ray emission is likely to be jet-dominated in nature.
\end{enumerate}

\acknowledgements

We wish to thank the anonymous referee for providing constructive comments. DAE gratefully acknowledges financial support for this work from NASA under grant number NNX11AD34G. This research has made use of data obtained from the Suzaku satellite, a collaborative mission between the space agencies of Japan (JAXA) and the USA (NASA). This research has also used the NASA/IPAC Extragalactic Database (NED) which is operated by the Jet Propulsion Laboratory, California Institute of Technology, under contract with NASA.


\clearpage

\clearpage

\begin{table}\footnotesize
\centering
\caption{Suzaku Observation Log}
\begin{tabular}{lllll}
\hline\hline
Date of observation    & Obs. ID & Instrument         & Screened exposure time & Net nuclear count rate (s$^{-1}$)    \\ \hline
2010 Dec 02               & 705039010         & XIS0       & 87 ks                 & 0.15$\pm$0.02 \\
                                   &                            & XIS1       & 87 ks                 & 0.21$\pm$0.02 \\
                                   &                            & XIS3       & 87 ks                 & 0.16$\pm$0.02 \\
                                   &                            & HXD/PIN    & 74 ks              & $(6.46\pm1.18)\times10^{-2}$ \\ \hline

\hline
\end{tabular}
\label{obslog}
\end{table}

\begin{table}\scriptsize
\centering
\caption{Best-fitting spectral parameters}
\begin{tabular}{lllllll}

\hline\hline

Model & Description    & $N_{\rm H}$ (cm$^{-2}$) & Power Law & Thermal 1 & Thermal 2 & $\chi^2$/dof \\ \hline

I          & PL                  &                                         & $\Gamma$=1.87$\pm$0.02                   &      &        &    1034/832 \\
           &                       &                                         & norm=(6.79$\pm$0.09)$\times10^{-4}$ &      &        &                    \\ \hline

II         & PL+TH1         &                                         & $\Gamma$=1.75$\pm$0.03                   & $kT$=$1.21^{+0.07}_{-0.10}$ keV          &        &    900/830 \\
           &                       &                                         & norm=(5.62$\pm$0.22)$\times10^{-4}$  & Z=0.35 (f)                                                &        &                    \\ 
           &                       &                                         &                                                                 & norm=(3.17$\pm$0.65)$\times10^{-4}$  &        &                    \\ \hline

III        & PL+TH1+TH2 &                                         & $\Gamma$=$1.62^{+0.07}_{-0.09}$                & $kT$=0.89$\pm$0.10 keV                                & $kT$=$2.63^{+0.69}_{-0.47}$ keV                         &    871/828	 \\
           &                       &                                         & norm=$(3.76^{+0.62}_{-0.79})\times10^{-4}$  & Z=0.35 (f)                                                         & Z=0.35 (f)                                                                &                    \\ 
           &                       &                                         &                                                                          & norm=$(2.04^{+0.48}_{-0.42})\times10^{-4}$  & norm=$(7.86^{+2.26}_{-2.36})\times10^{-4}$         &                    \\ \hline

IV        & ABS(PL)+TH1+TH2 & $(7.68^{+1.26}_{-2.18})\times10^{20}$ & $\Gamma$=$1.82^{+0.04}_{-0.05}$                & $kT$=$0.80\pm0.15$ keV                   & $kT$=$1.42^{+0.45}_{-0.20}$ keV                         &    838/827 \\
           &                                 &                                                               & norm=$(6.17^{+0.38}_{-0.54})\times10^{-4}$  & Z=0.35 (f)                                                         & Z=0.35 (f)                                                                &                    \\ 
           &                                 &                                                               &                                                                          & norm=$(1.27^{+0.38}_{-0.63})\times10^{-4}$  & norm=$(2.22^{+1.04}_{-0.66})\times10^{-4}$         &                    \\ 

\hline
\end{tabular}
\label{results}
\footnotetext{Col. (1): Model number. Col. (2): Description of spectrum (Abs=Neutral absorption, PL=Power Law, TH=Collisionally ionized plasma model). Col. (3): Intrinsic neutral hydrogen column density. Galactic absorption has also been applied. Col. (4): Power Law parameters. Normalization is quoted at 1 keV in units of ph~keV$^{-1}$~cm$^{-2}$~s$^{-1}$. Col. (5): Parameters of first collisionally ionized plasma model. Col. (6): Parameters of second collisionally ionized plasma model. Col. (7): Value of $\chi^{2}$ and degrees of freedom.}
\end{table}

\end{document}